\documentclass[showpacs,superscriptaddress,aps,twocolumn,prl]{revtex4-1} 

\usepackage[USenglish]{babel}
\usepackage{graphicx,xcolor}

\usepackage{amsmath, bbm}
\usepackage{mathtools}
\usepackage{hyperref}
\usepackage{epstopdf}
\usepackage{color}
\usepackage{notes2bib}
\usepackage{siunitx}

\begin{document}

\title{Nanoscopic charge fluctuations in a gallium phosphide waveguide\\ measured by single molecules}
 \author{Alexey Shkarin*}
\affiliation{Max Planck Institute for the Science of Light, D-91058 Erlangen, Germany}
 \author{Dominik Rattenbacher*}
\affiliation{Max Planck Institute for the Science of Light, D-91058 Erlangen, Germany}
\author{Jan Renger}
\affiliation{Max Planck Institute for the Science of Light, D-91058 Erlangen, Germany}
\author{Simon H\"onl}
\affiliation{IBM Research Europe, S{\"{a}}umerstrasse 4, CH-8803 R{\"{u}}schlikon, Switzerland}
\author{Tobias Utikal}
\affiliation{Max Planck Institute for the Science of Light, D-91058 Erlangen, Germany}
\author{Paul Seidler}
\affiliation{IBM Research Europe, S{\"{a}}umerstrasse 4, CH-8803 R{\"{u}}schlikon, Switzerland}
\author{Stephan G\"otzinger}
\affiliation{Department of Physics, Friedrich Alexander University Erlangen-Nuremberg, D-91058 Erlangen, Germany}
\affiliation{Max Planck Institute for the Science of Light, D-91058 Erlangen, Germany}
\affiliation{Graduate School in Advanced Optical Technologies (SAOT), Friedrich Alexander
University Erlangen-Nuremberg, D-91052 Erlangen, Germany}
\author{Vahid Sandoghdar}
\affiliation{Max Planck Institute for the Science of Light, D-91058 Erlangen, Germany}
\affiliation{Department of Physics, Friedrich Alexander University Erlangen-Nuremberg, D-91058 Erlangen, Germany}

\begin{abstract}
We present efficient coupling of single organic molecules to a gallium phosphide subwavelength waveguide (nanoguide). By examining and correlating the temporal dynamics of various single-molecule resonances at different locations along the nanoguide, we reveal light-induced fluctuations of their Stark shifts. Our observations are consistent with the predictions of a simple model based on the optical activation of a small number of charges in the GaP nanostructure. \\\\\\ * A.S. and D.R. contributed equally to this work.

\end{abstract}

\maketitle

One of the promising platforms for future quantum technologies is based on integrated nanophotonics, where a large number of quantum emitters and quantum states of light are efficiently interconnected via a labyrinth of subwavelength waveguides (nanoguides) and other nano-optical elements \cite{Wang2020,Kim20,Elshaari2020,Sandoghdar2020,Tuerschmann2019}. The optimal choice of materials for achieving this goal has not been settled yet, but semiconductors (e.g., GaAs) \cite{Thyrrestrup2018, Hallett2018, Grim2019} and diamond \cite{Atature2018} architectures have made significant progress. In most cases, quantum emitters have been situated inside the host material of the nanoguides, e.g., AlGaAs quantum dots in GaAs or color centers in diamond. To decouple the choices of ideal quantum emitters, appropriate guiding architectures, and other nano-optical elements from each other, there is a large effort towards hybrid solutions \cite{Akimov07,Tuerschmann2017,Lombardi2018,Grandi2019,Hail19,Froech20,Elshaari2020,Kim20}. 

Guiding light on a chip is best realized in a material with high refractive index, $n$. Common choices for waveguides in the visible and near-infrared regimes have been silicon nitride \cite{Lombardi2018,Wan2020,Froech20} or titanium dioxide \cite{Tuerschmann2017,Rattenbacher2019} with $n\sim2-2.5$. Semiconductors such as Si and GaAs offer considerably larger $n$ but at the expense of strong absorption for wavelengths smaller than \SI{800}{nm}. Gallium phosophide (GaP) presents a very attractive alternative with $n\gtrsim3$ and a cutoff wavelength below \SI{600}{nm} \cite{Barclay2009,Wolters2010,Gould2016,Wilson2020}, and indeed, recent efforts have established fabrication of high-quality nanophotonic chips from this material on $\mathrm{SiO_2}$ \cite{Hoenl2018, Schneider2018}. In this work, we show efficient coupling of single organic molecules to a GaP nanoguide. Furthermore, we investigate light-induced charge fluctuations in the nanoguide by analyzing the spatio-temporal features of single-molecule Stark shifts.

\begin{figure}[t!]
	\includegraphics[width=\columnwidth]{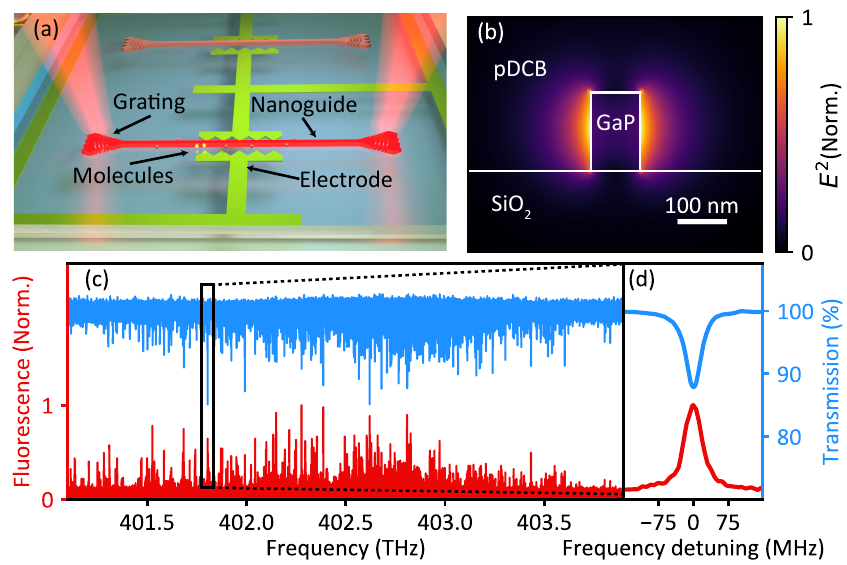}
	\caption{(a) Schematic of a nanoguide interfaced with two electrodes. (b) Simulated intensity profile of the TE mode in the nanoguide. (c) Extinction (blue) and fluorescence (red) spectra of hundreds of molecules. (d) Extinction and fluorescence spectra of one single molecule. 
	\label{fig:overview}}
\end{figure}

\textit{Efficient coupling of molecules to a GaP nanoguide.} The core of our experimental platform is a \SI{50}{\micro\meter} long GaP nanoguide fabricated on a \SI{2}{\micro\meter} thick $\mathrm{SiO_2}$ layer on a silicon substrate \cite{Hoenl2018, Schneider2018}. The nanoguide has a cross-section of $100\,\rm{nm}\times160\,\rm{nm}$, is terminated on both sides by grating couplers and is decorated by sawtooth-shaped gold microelectrodes placed \SI{2}{\micro\meter} from the nanoguide (see Fig.\,\ref{fig:overview}(a, b)). The nanoguide supports two propagating modes of orthogonal polarization states, which have the majority of their energy concentrated in the evanescent field (see Fig.\,\ref{fig:overview}(b)). Fabrication details are presented in earlier works \cite{Hoenl2018, Schneider2018,Wilson2020}.
The nanostructures are covered by a para-dichlorobenzene (pDCB) crystal confined in a nanochannel as described in previous publications \cite{Gmeiner2016,Tuerschmann2017,Rattenbacher2019}. By doping the pDCB crystal with dibenzoterrylene (DBT) at a concentration of $2\times10^{-4}$ mol/mol, we realize a random distribution of the latter. At a temperature below \SI{4}{K}, DBT molecules possess strong zero-phonon lines with a lifetime-limited linewidth of \SI{30}{MHz} at a vacuum transition wavelength of \SI{745}{nm} (frequency of \SI{402}{THz}) \cite{Nicolet2007,Verhart2014}. Owing to variations in their nanoscopic crystal environment \cite{Gmeiner2016}, the resonance frequencies of individual molecules span a range of about \SI{4}{nm} (\SI{2}{THz}). The transverse electric (TE) mode of the nanoguide (see Fig.\,\ref{fig:overview}(b)) is excited by continuous-wave Ti:Sapphire laser beams (linewidth\,$<$\,\SI{1}{MHz}), and the outcoupled radiation is detected on an avalanche photodiode. The blue curve in Fig.\,\ref{fig:overview}(c) displays the extinction spectra of several hundred molecules measured in transmission, while the red spectrum reports on the corresponding fluorescence via the vibrational levels of the ground state and phonon wings. In Fig.\,\ref{fig:overview}(d), we plot a closeup for a single-molecule extinction dip of 13\%, comparable with the coupling efficiency achieved for illumination in a diffraction-limited focus spot \cite{Wrigge2008}.
 
\begin{figure}[t!]
	\includegraphics[width=\columnwidth]{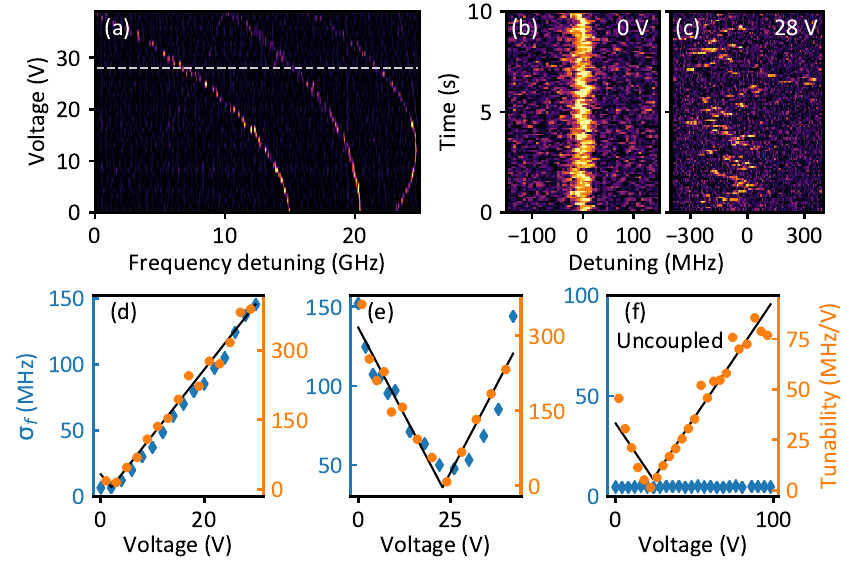}
	\caption{(a) Fluorescence signals of three molecules as a function of laser frequency detuning and applied external voltage. The dashed line points to broad instable resonances at high voltages. (b, c) Temporal spectral dynamics of a single nanoguide-coupled molecule (M1) studied under two different applied DC voltages of \SI{0}{V} and \SI{28}{V}. Individual spectra were recorded at the rate of 10 scans/s. The color scheme in (a-c) is chosen to highlight the molecular resonance frequency. (d, e) Frequency fluctuations $\sigma_f$ (blue, diamond) and absolute tunability $\frac{\partial f}{\partial V}$ (orange, circle) as a function of applied voltage for two nanoguide-coupled molecules. The molecule in (d) was also M1. (f) Same as (d, e) but for a molecule far from the nanoguide. Black lines are linear fits to the absolute value of tunability. 
		\label{fig:tuning}}
\end{figure}

\textit{Monitoring the local electric field.} The molecule DBT has inversion symmetry and is thus expected to undergo a quadratic Stark shift. Figure\,\ref{fig:tuning}(a) illustrates the parabolic spectra of three nanoguide-coupled molecules as a function of the voltage applied to the nearby electrodes. In some cases, local strain in the matrix leads to the addition of a linear Stark effect manifested by an offset in the turning points of the parabolas \cite{ORRIT1992,Moradi2019}. A more noteworthy observation in these spectra is, however, that the resonances appear broader and less stable at larger slopes of the Stark tuning profile (see e.g., the spectra across the dashed line in Fig.\,\ref{fig:tuning}(a)). We point out that this effect is consistently observed for all nanoguide-coupled molecules on all studied GaP samples. On the contrary, we have verified that molecules at about \SI{10}{\micro\meter} away from the nanoguide are highly stable and have a lifetime-limited linewidth regardless of the applied Stark shift. For the latter measurements, we shined a focused laser beam normal to the plane of the chip.

We, thus, attribute the origin of the spectral instabilities for nanoguide-coupled molecules to electric field fluctuations originating from the GaP nanostructure. As we show in the following, the effect at hand is very different from previous reports of electric field-induced spectral instability and broadening of organic molecules, which were caused by two-level tunneling systems \cite{Phillips1972,Heuer1993,Maier1995,Segura2001,Bauer2003,Gerhardt2009} or electro-mechanical oscillations \cite{Tian2014} in the host organic matrix. 

To investigate the temporal dynamics of the spectral instabilities more quantitatively, we increased the frequency scan rate from \SI{400}{MHz/s} in Fig.\,\ref{fig:tuning}(a) to \SI{10}{GHz/s}. This was sufficient to arrive at lifetime-limited Lorentzian resonances in individual frequency sweeps and, thus, resolve the wandering of their center frequencies over time. To characterize Stark spectra such as those shown in Fig.\,\ref{fig:tuning}(a), we recorded a large number of individual scans.  In this manner, we obtained a robust mean value ($f$) for the molecular resonance frequency and root mean square (RMS) frequency fluctuations ($\sigma_f$) at a given electrode voltage ($V$). By repeating this procedure for many applied voltages, we established the frequency tunability defined as $\frac{\partial f}{\partial V}$. Figures\,\ref{fig:tuning}(b) and \ref{fig:tuning}(c) display examples of spectral trajectories of a molecule, which we name M1, at two applied voltages.
Figure\,\ref{fig:tuning}(d-f) shows $\sigma_f$ and the tunability for three exemplary molecules. For M1, $\frac{\partial f}{\partial V}$ (orange) increases linearly with $V$, as shown in Fig.\,\ref{fig:tuning}(d), implying a quadratic Stark behavior. Interestingly, $\sigma_f$ (blue) also grows proportionally with $\frac{\partial f}{\partial V}$. Figure\,\ref{fig:tuning}(e) presents the case of a second nanoguide-coupled molecule that behaves similarly although its turning point ($\frac{\partial f}{\partial V}=0$) is shifted to a voltage of about \SI{25}{V}. As a control experiment, in Fig.\,\ref{fig:tuning}(f) we report on a molecule far away from the nanoguide. It is evident that, in this case, there is no correlation between $\frac{\partial f}{\partial V}$ and $\sigma_f$, which stays at 4\,MHz, given by the fit error. 

Assuming that the field fluctuations originate from the nanoguide, we express the total Stark shift experienced by the molecule as $\delta f\propto \lvert\mathbf{E}_{\rm cr}+\mathbf{E}_{\rm el}+\mathbf{E}_{\rm ng}\rvert^2$, where $\mathbf{E}_{\rm cr}$, $\mathbf{E}_{\rm el}$ and $\mathbf{E}_{\rm ng}$ denote the fields created by the residual strain in the crystal, the electrodes, and the nanoguide, respectively. A cross term in this expression leads to the amplification of a small fluctuating $\mathbf{E}_{\rm ng}$ by a large constant $\mathbf{E}_{\rm el}+\mathbf{E}_{\rm cr}$. The former quantity causes $\sigma_f$ while the latter dictates $\frac{\partial f}{\partial V}$ such that $\sigma_f$ becomes proportional to $\frac{\partial f}{\partial V}$, as observed in our measurements. The measured data in Fig.\,\ref{fig:tuning}(e, d) and our knowledge of the electrode and nanoguide geometries let us deduce the RMS nanoguide field fluctuations to be around $\SI{40}{kV/m}$ at the position of the molecule. This corresponds to the field generated by a single electron at a distance of \SI{100}{nm}, which is comparable to the typical separation of the evanescently-coupled nanoguide-molecule system.

\begin{figure}[t!]
	\includegraphics[width=\columnwidth]{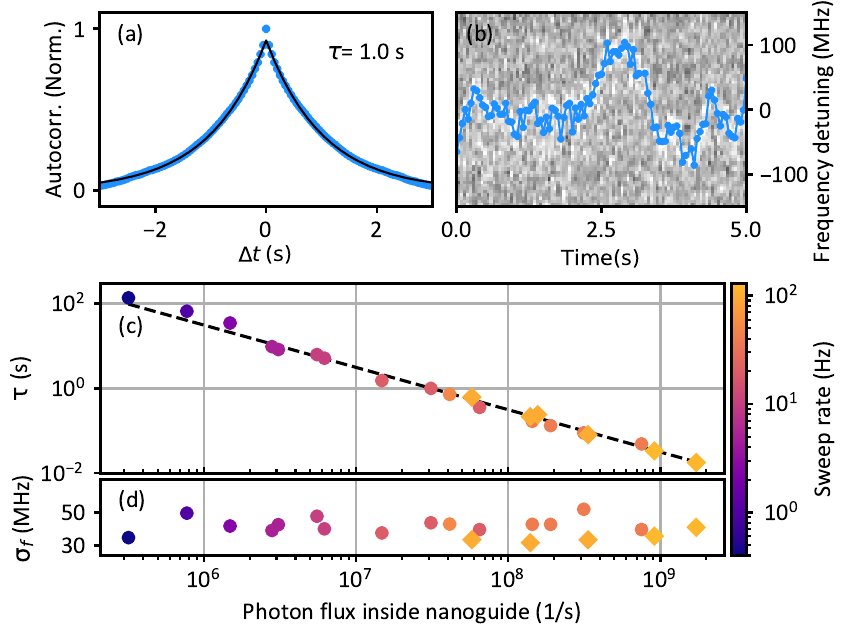}
	\caption{(a) Autocorrelation function of the frequency trace of molecule M1 normalized to $\mathrm{C}_{11}(0)$, measured at 10\,V and an optical power of $3\times10^7$ photons/s. Black line shows an exponential fit function, yielding an autocorrelation time of $\tau=\SI{1.0}{s}$. (b) Nanoguide transmission for repeated scans of the laser frequency. The blue symbols highlight the fitted resonance frequencies for each sweep. (c) Dependence of $\tau$ on the optical power ($P$) in the nanoguide. The power was referenced to the photon flux inside the nanoguide by analysing the saturation behavior of the molecules \cite{Tuerschmann2017}. Dots (diamonds) denote results obtained through extinction (fluorescence) measurements. The color-coding indicates the sweep rate used to record the frequency traces. The straight line shows a fit to $\tau\propto P^{-1}$. (d) Magnitude of the center frequency fluctuations.
		\label{fig:dynamics_ac}}
\end{figure}

\textit{Temporal field correlations.} The ability to scan faster than typical electric field fluctuations allows us to determine the correlation function $\mathrm{C}_{jk}(\Delta t)$ between the frequency fluctuations of molecules $j$ and $k$. Figure\,\ref{fig:dynamics_ac}(a) plots the autocorrelation function $\mathrm{C}_{11}$ computed for molecule M1 from 12,000 individual sweeps, 100 of which are displayed in Fig.\,\ref{fig:dynamics_ac}(b). By fitting an exponential function to $\mathrm{C}_{11}$, we extract the autocorrelation time $\tau=\SI{1.0}{s}$.

Next, we studied the dependence of $\tau$ and $\sigma_f$ on the optical power ($P$) in the nanoguide. Figure\,\ref{fig:dynamics_ac}(c, d) displays the outcome of measurements for one molecule, indicating an inverse proportionality relation between $\tau$ and $P$ over four orders of magnitude. This robust dependency clearly demonstrates that the fluctuations are photo-induced. We checked that the observed phenomenon is not caused by direct interaction of light with the molecule. To do this, we performed measurements, where up to 90\% of the optical power inside the nanoguide was provided by a second laser beam detuned by 2\,THz from the molecule and verified that the fluctuation rate ($1/\tau$) scaled with the total laser power in the nanoguide. Furthermore, we found that $\sigma_f$ remains independent of $P$ within our measurement precision (see Fig.\,\ref{fig:dynamics_ac}(d)). This indicates that while the external illumination power dictates the spectral noise dynamics, it does not influence its amplitude.

\begin{figure}[b!]
	\includegraphics[width=\columnwidth]{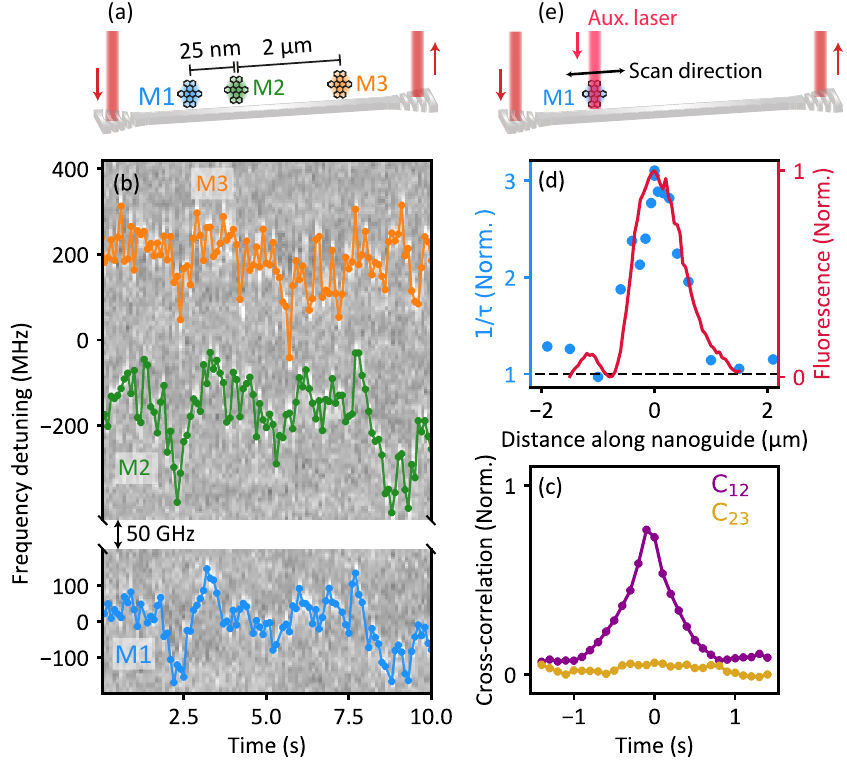}
	\caption{(a) Sketch of the arrangement of molecules labeled M1, M2 and M3 along the nanoguide. (b) Examples of transmission spectra obtained simultaneously at two frequency intervals separated by about \SI{50}{GHz}. Colored lines indicate the evolution of the resonance frequencies. (c) Cross-correlation functions of M1-M2 ($\mathrm{C}_{12}$, magenta) and M2-M3 ($\mathrm{C}_{23}$, mustard). (d) Frequency fluctuation rate $1/\tau$ (blue) as a function of the displacement of an auxiliary focused laser beam along the nanoguide over a molecule; see (e).  The rate is normalized to that in the absence of the auxiliary excitation. The red curve shows the intensity profile of the excitation laser's focus spot mapped by the fluorescence of the molecule. (e) Sketch of the arrangement when a second laser beam is shone in the direction normal to the plane of the chip and its focus spot is scanned along the nanoguide.
		\label{fig:crosscorrelation}}
\end{figure}

\textit{Spatial field correlations.} We now investigate the spatial properties of the observed electric field noise by simultaneous measurement of several molecules. Here, we coupled two independent laser beams into the nanoguide and performed time-multiplexed alternate frequency sweeps in two different frequency regions. In Fig.\,\ref{fig:crosscorrelation}, we present an example, where the frequency fluctuations of three molecules (M1, M2 and M3; see Fig.\,\ref{fig:crosscorrelation}(a)) were examined. The frequency variations and the cross-correlation functions $\mathrm{C}_{12}, \mathrm{C}_{23}$ displayed in Figs.\,\ref{fig:crosscorrelation}(b) and \ref{fig:crosscorrelation}(c), respectively, reveal that the spectral fluctuations in M1 and M2 are correlated, but M3 behaves independently. Considering the error in extracting resonance frequencies, we infer almost perfect correlation between M1 and M2. By determining the positions of each molecule through localization microscopy, we revealed that, indeed, M1 and M2 were separated by only $26\pm5$\,nm, while M3 was about \SI{2}{\micro\meter} away from them. For another correlated pair separated by about \SI{80}{nm} we saw that the peak spectral cross-correlation drops to $\sim30\%$, verifying that the field fluctuations are very local. 

To provide further evidence for the local character of the field fluctuations, we also conducted single-molecule measurements under illumination by an auxiliary laser beam that was frequency detuned by 6\,GHz. Here, we scanned the focal spot of this second beam along the nanoguide across the molecule position (see Fig.\,\ref{fig:crosscorrelation}(d,e)) and recorded changes in $1/\tau$. The blue symbols in Fig.\,\ref{fig:crosscorrelation}(d) confirm that $1/\tau$ follows the intensity profile of the auxiliary light beam shown by the red curve (full width at half-maximum \SI{1}{\micro\meter}) 

\textit{Theoretical modelling.} Regardless of the details of the process at work, electric field fluctuations generated by the nanoguide can be attributed to charge fluctuations. Redistribution of charges in semiconductors can be caused by many effects such as trapped or wandering charges \cite{Mueller2005,Sallen2011,Bardoux2006, Wolters2013,Thoma2016,Liu2018}, impurities \cite{Hauck2014}, or ligand rearrangements \cite{Fernee2012,Beyler2013} and have also been shown to be driven by light with energy below the bandgap \cite{SAOS}. We consider a simple model in which electric field fluctuations result from the rearrangement of randomly distributed point charges with density $n_{\rm q}$ in the GaP nanoguide. We assume that each charge stays in the vicinity of its original position, but it experiences an average displacement $d$ upon scattering one photon. In other words, while the field fluctuation rate scales with the optical power $P$, $n_{\rm q}$ and $d$ remain independent of it. It can then be shown that $\sigma_f \propto d\sqrt{n_{\rm q}}$. Furthermore, the randomness of the jumps and the quadratic fall-off of the Coulomb field lead to a short correlation length, approximately equal to the molecule-waveguide distance.

\begin{figure}[bt!]
	\includegraphics[width=\columnwidth]{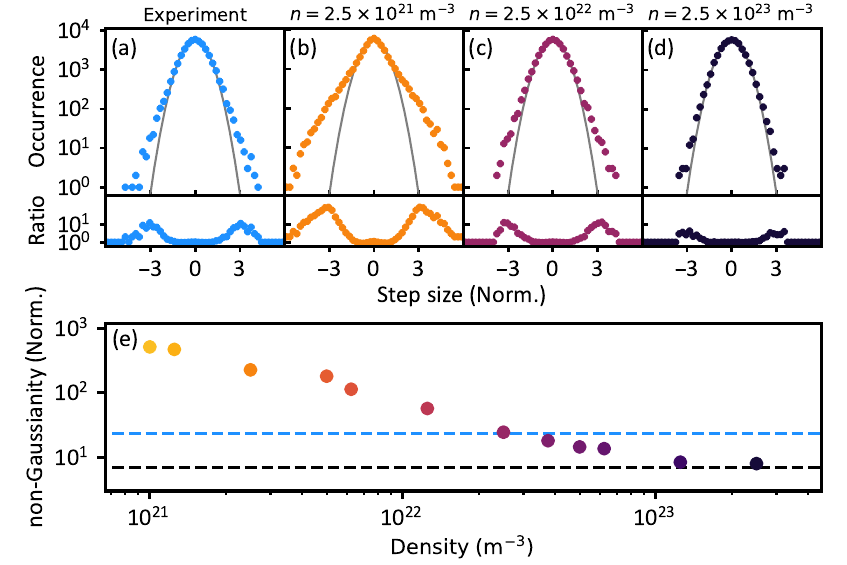}
	\caption{Upper plots: Histograms of the step size distributions for the the experimental data (a) and three synthetic data sets at different densities $n$ (b-d). Grey curves denote fits to a Gaussian distribution. Lower plots: Ratio of the occurrences to their corresponding Gaussian functions after adding a regularization offset of 1. This offset  reduces the contribution of histogram bins with small number of counts, whose exact values are dominated by statistical noise. (e) The symbols show the dependence of the non-Gaussianity parameter $\eta$ (see text) on the simulated charge density used. The blue horizontal line corresponds to the experimental data, and the black line signifies the baseline of our analysis caused by fit errors.
		\label{fig:step_distribution}}
\end{figure}

In the limit of large $n_{\rm q}$, one expects a Gaussian statistics for the frequency steps. The upper panel of Fig.\,\ref{fig:step_distribution}(a) shows that the histogram of the measured data clearly deviates from a Gaussian distribution shown in grey. To compare the statistical properties of the frequency fluctuations with the predictions of our model, we simulated time traces of the molecular resonance frequencies for different $n_{\rm q}$ values and used the same analysis procedure applied to the experimental data to generate a histogram of the frequency steps.  The upper panel of Fig.\,\ref{fig:step_distribution}(b-d) displays the outcome for three synthetic data sets. It turns out that at the lowest considered density, corresponding to the effect of only $\sim5$ charges, the molecular resonance experiences jump-like frequency shifts, causing a non-Gaussian frequency step-size distribution. 
 
To quantify the deviations from a Gaussian distribution, we consider the ratio of the occurrence to the corresponding Gaussian fit distribution as plotted in the lower panel of Fig.\,\ref{fig:step_distribution}(a-d). Next, as presented in Fig.\,\ref{fig:step_distribution}(e), we introduce a measure of non-Gaussianity ($\eta$), defined as the area under this curve above one. The experimentally measured $\eta$ marked by the blue line corresponds to a density of charges around $n_{\rm q}=\SI{2.5d22}{m^{-3}}$, or equivalently to about one charge per $(\SI{35}{nm})^{3}$. Although this analysis only established an estimate, the result is in agreement with the typical density of defects, impurities or charge traps seen in such systems \cite{Hauck2014} and implies that an average of 50 charges govern the frequency fluctuations of our molecules. Given this density and the experimentally measured magnitude of the frequency fluctuations, we can estimate $d \approx\SI{20}{nm}$. We also note that owing to the small width of the nanoguide, surface and volume charges lead to similar results.

\textit{Conclusions.} We have used single molecules as nanometer-sized probes for investigating the spatio-temporal behavior of a low number of charges activated in GaP nanoguides. The small size, excellent spectral properties, large achievable concentrations and the inhomogeneous distribution of their resonance frequencies make organic molecules a promising tool for ultrasensitive characterization of nanoscopic charge dynamics in a range of systems such as single electron transistors, quantum dots or superconductors \cite{Rezai2018,Caruge2001,Faure2007,Vamivakas2011,Arnold2014, Faez2014}. Our findings also advance the use of GaP as a platform for integrated quantum photonics \cite{Wilson2020,Sandoghdar2020,Wang2020,Kim20}. The observed light-induced field fluctuations are small and slow enough to be tolerated or eliminated by more sophisticated fabrication schemes \cite{Guha2017,Liu2018}. However, even in their current form, the estimated density of charges and their light absorption probability signified by the slope of Fig.\,\ref{fig:dynamics_ac}(c) point to a loss coefficient of about \SI{0.5}{dB/cm}, which would allow for resonator quality factors in the order of $10^6$.

\textit{Acknowledgment.} We acknowledge financial support by the Max Planck Society as well as the RouTe Project (13N14839) through the Federal Ministry of Education and Research (BMBF) and the European Union's Horizon 2020 Program for Research and Innovation under grants 722923 (Marie Curie H2020-ETN OMT) and 732894 (FET Proactive HOT). A.S. acknowledges support through an Alexander von Humboldt fellowship.

%


\end{document}